%% file: template.tex
\documentclass[a4paper]{article}

\usepackage{INTERSPEECH2022}
\usepackage[OT1]{fontenc} % so that section titles are in boldface
\usepackage{xcolor}
\usepackage{url}
\usepackage{hyperref}
\usepackage{multirow}
\usepackage{amsmath} % for argmax

\usepackage[ruled]{algorithm2e}

\newcommand{\eqdef}{\overset{def}{=}}

\title{Pruned RNN-T for fast, memory-efficient ASR training}
\name{
Fangjun Kuang,
Liyong Guo,
Wei Kang,\\
Long Lin,
Mingshuang Luo,
Zengwei Yao,
Daniel Povey}
\address{Xiaomi Corp., Beijing, China}
\email{\{kuangfangjun, guoliyong, kangwei1, linlong, luomingshuang, yaozengwei, dpovey\}@xiaomi.com}

\begin{document}

\maketitle
\input{abstract}

\noindent\textbf{Index Terms}: speech recognition, transducer, end-to-end, limited memory

\input{intro}
\input{motivations}
\input{pruned-rnnt}
\input{experiments}
\input{results}
\input{conclusion}

\bibliographystyle{IEEEtran}

\bibliography{mybib}

\end{document}

%% file: abstract.tex
\begin{abstract}
  The RNN-Transducer (RNN-T) framework for speech recognition has been growing
  in popularity, particularly for deployed real-time ASR systems, because it
  combines high accuracy with naturally streaming recognition.
  One of the drawbacks of RNN-T is that its loss function is relatively slow to
  compute, and can use a lot of memory.  Excessive GPU memory usage can make it
  impractical to use RNN-T loss in cases where the vocabulary size is large:
  for example, for Chinese character-based ASR.

  We introduce a method for faster and more memory-efficient RNN-T loss computation.
  We first obtain pruning bounds for the RNN-T recursion using a simple joiner
  network that is linear in the encoder and decoder embeddings; we can evaluate
  this without using much memory.  We then use those pruning bounds to evaluate
  the full, non-linear joiner network. The code is open-sourced and publicly
  available.
\end{abstract}

%% file: intro.tex
\section{Introduction}

End-to-End (E2E) models have been growing in popularity in the field of automatic
speech recognition (ASR). Unlike conventional ASR, which contains an acoustic
model (AM) and a language model (LM) that are usually trained separately,
E2E models use a single neural network model to predict words or graphemes
directly from acoustic waveforms, which simplifies both training and decoding.

Three popular E2E models are connectionist temporal classification (CTC)
models~\cite{graves2006connectionist}, attention-based
models~\cite{vaswani2017attention}, and RNN-T
models~\cite{graves2012sequence}.
RNN-T models are naturally streaming and can be decoded frame
synchronously.  On one hand, it does not require the full context to predict the
next token, as is required by attention-based models. On the other hand, there
are no assumptions about frame independence given acoustic inputs that exist in
CTC models. As a consequence, RNN-T models are very attractive in industry
areas~\cite{he2019streaming, botros2021tied}.

The output of the RNN-T model~\cite{graves2012sequence} is
usually a 4-D tensor of shape $(N, T, U, V)$, where $N$ is the batch size,
$T$ is
the output length of the transcription network, $U$ is the output length of
the prediction network, and $V$ is the vocabulary size. The output contains
the probability distribution over all tokens in the utterance at each time
step, which requires a lot of memory in training.  For large vocabulary sizes
(e.g. Chinese characters), this can severely limit the batch size and slow
down training.

There are various efforts to reduce memory usage in RNN-T training.
One technique is to remove paddings when combining
the outputs from the prediction network and the transcription
network~\cite{li2019improving}. Function merging~\cite{li2019improving} is also
used to reduce memory consumption by computing gradient with respect to the logits
directly. Another method is to use half-precision for training~\cite{zhang2021benchmarking}
at the cost of degradation in WER.

In this paper, instead of generating a probability distribution over all the
tokens $U$ at each time step, we propose pruned RNN-T, which limits the range
of tokens at each time step from $U$ to $S$, where
$S \ll U$. Therefore, the output shape becomes $(N, T, S, V)$, leading to
less memory usage and faster training. We show that using
pruned RNN-T for training can not only reduce memory consumption but can also
achieve faster training, without performance degradation in WER.
Furthermore, the above-mentioned techniques can be used together with
pruned RNN-T to further reduce memory usage and accelerate training.

In a slight abuse of notation, when we refer to RNN-T in this paper we are
speaking of the RNN-T loss itself, which is more properly speaking the
transducer loss.  In our experiments we use a Conformer
encoder~\cite{anmol2020conformer}, not a recurrent encoder; and the decoder is
stateless~\cite{ghodsi2020rnntstateless} rather than recurrent.

The code for this work is open-sourced and publicly
available\footnote{\url{https://github.com/danpovey/fast_rnnt}
and
\url{https://github.com/k2-fsa/k2}
}.

The remainder of the paper is structured as follows. In
Section~\ref{sec:motivations}, we briefly describe the standard RNN-T
and identify the reason for its high demand for memory in training, motivating
us to propose pruned RNN-T in Section~\ref{sec:pruned-rnnt} to reduce memory
consumption and to accelerate training. Section~\ref{sec:experiments} gives
the experiment setup for benchmarking different implementations of RNN-T loss
and applying pruned RNN-T in ASR training. The results
are given in Section~\ref{sec:results}.
Finally, we conclude the paper in Section~\ref{sec:conclusions}.

%% file: motivations.tex
\section{Motivations}
\label{sec:motivations}

Assume the acoustic input has been parameterized into a sequence of~$T$
feature frames $\mathbf{x}=\{x_t\}_{t=0}^{T{-}1}$.
Also assume the transcript has been tokenized into a sequence of $U$ tokens
$\mathbf{y} = \{0 \leq y_u < V | u=0, 1, 2, \ldots, U \}$,
where $V$ is the vocabulary size and token ID 0 is the blank token $\varnothing$;
we have $\mathbf{y}_0 = 0$ as a beginning-of-sentence token, with the
remaining positions corresponding to ``real'' words (there is no end-of-sentence
token).

There are 3 components in the standard transducer model~\cite{graves2012sequence}:
An encoder (a.k.a transcription network), a decoder (a.k.a prediction network),
and a joiner (a.k.a  joint network).

The encoder network functions as an acoustic model, transforming
acoustic frames into a high-level representation. The encoder output is a
2-D tensor $\mathbf{X}$ with shape $(T, E)$\footnote{We assume the batch size is 1.},
where $E$ is the output dimension of the encoder.
The decoder network is similar to a language model that tries to predict
a distribution by conditioning on the last non-blank token. The output of the
decoder is a 2-D tensor $\mathbf{Y}$ of shape $(U+1, D)$, where $D$ is the output
dimension of the decoder. The joiner combines the outputs from the encoder
network and the decoder network and produces a log-probability
distribution $\mathbf{L}$ of shape $(T, U{+}1, V)$, where $L(t,u,v)$ is the
log-probability for the token $v$ to appear at position $t$, given $y_{0..u}$.

\begin{figure}[t]
\centering
\begin{tabular}{cc}
\includegraphics[width=0.24\textwidth,page=1]{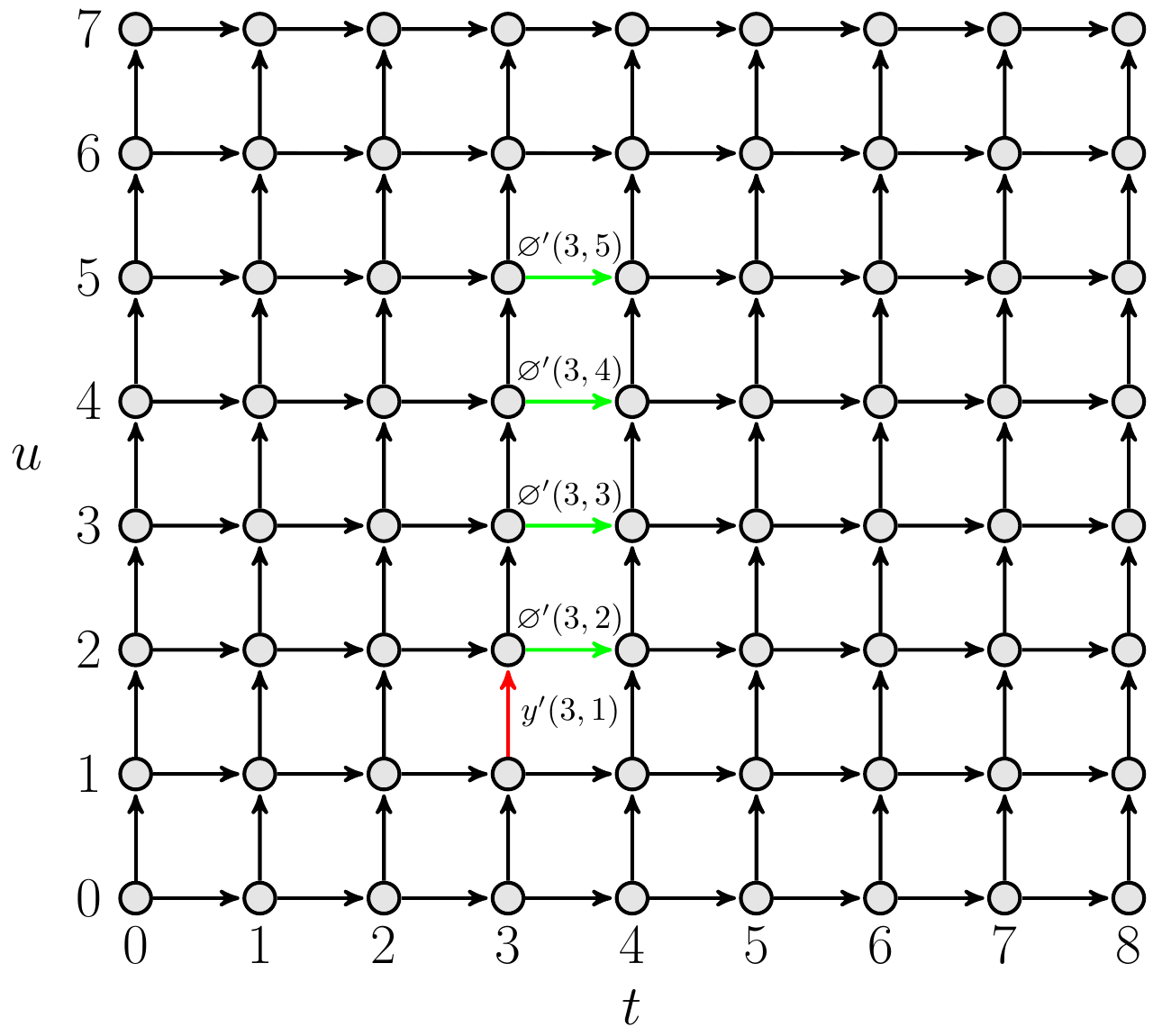}&
\includegraphics[width=0.24\textwidth,page=2]{tikz/pic.pdf}\\
\textbf{(a)} & \textbf{(b)}
\end{tabular}
  \caption{Output log-probability lattice defined by the joiner output $\mathbf{L}$
in RNN-T. The vertical transition leaving node $(t,u)$ has
the log-probability $y(t,u)$, while the horizontal leaving node $(t,u)$
has the log-probability $\varnothing(t,u)$.
Log-probabilities for transitions
that are not shown in the figure are set to minus infinity.
\textbf{(a)} Lattice for the standard RNN-T. \textbf{(b)} Lattice for the pruned RNN-T.
}
   \label{fig:lattice}
  \vspace{-2mm}
\end{figure}

Similar to~\cite{graves2012sequence} (but in log-space, and with zero-based $t$ index), we define
\begin{align}
y(t, u) &= L(t, u, y_{u+1}) \\
\varnothing(t, u) &= L(t, u, \varnothing)
\end{align}
where $y(t,u)$ is the log-probability of the vertical transition leaving the
node at position $(t,u)$ in Figure~\ref{fig:lattice}(a),
while $\varnothing(t,u)$ is the log-probability of the horizontal transition
leaving the node at position $(t,u)$ in Figure~\ref{fig:lattice}(a).

It usually uses the forward-backward algorithm~\cite{graves2012sequence} to
compute the RNN-T loss. Let the forward variable $\alpha(t,u)$  be the
log-probability outputting $y_{0..u}$ after seen $x_{0..t}$.
It can be computed recursively using~\eqref{eq:standard-forward}
\begin{align}
\alpha(t,u) = \mathrm{LogAdd}&\left( \alpha(t-1, u)+ \varnothing(t-1, u) \right.,\nonumber\\
&\left. \alpha(t, u-1) +  y(t, u-1) \right) \label{eq:standard-forward}
)
\end{align}
where $\mathrm{LogAdd}$ is defined as:
\begin{equation}
\mathrm{LogAdd}(x,y) = \log(e^x + e^y)
\end{equation}

$\alpha(0,0)$ is initialized to 0 and the total log-likelihood of the sequence
is $\alpha(T{-}1,U) + \varnothing(T{-}1, U)$.

The RNN-T loss computation can be quite memory- and compute-intensive because it
has to compute $y(t,u)$ and $\varnothing(t,u)$ for all $t \le T$ and $u\le U$,
so the output shape of the joint network has to be $(N, T, U, V)$ if the batch size
is $N$.  This is much larger than the output required by CTC~\cite{graves2006connectionist} and
attention-based~\cite{vaswani2017attention}
models which involves shapes like $(N, T, V)$ or $(N, U, V)$.

\begin{figure}[t]
\centering
\begin{tabular}{c}
\includegraphics[width=0.45\textwidth]{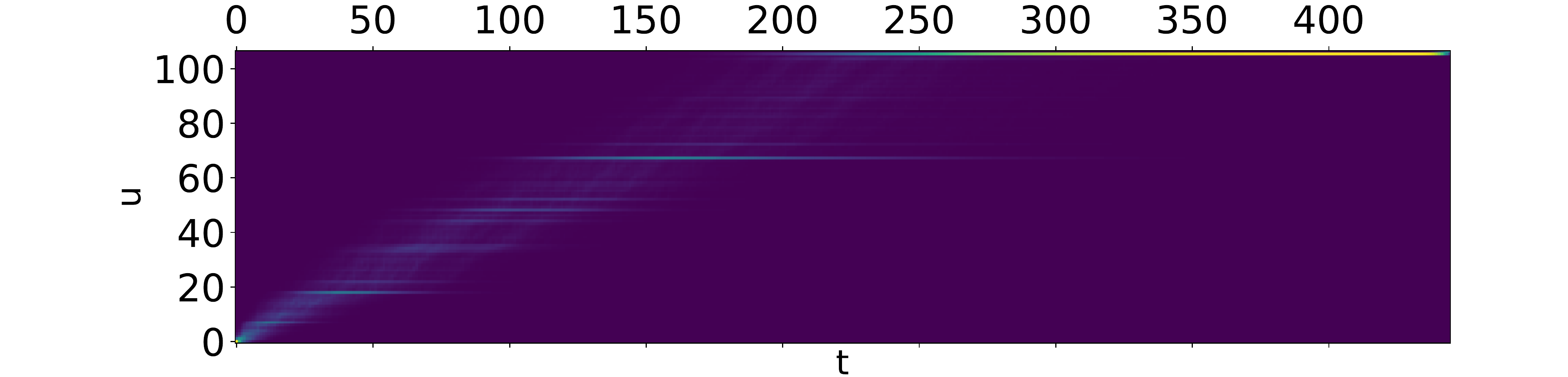}\\
\textbf{(a)}\\
\includegraphics[width=0.45\textwidth]{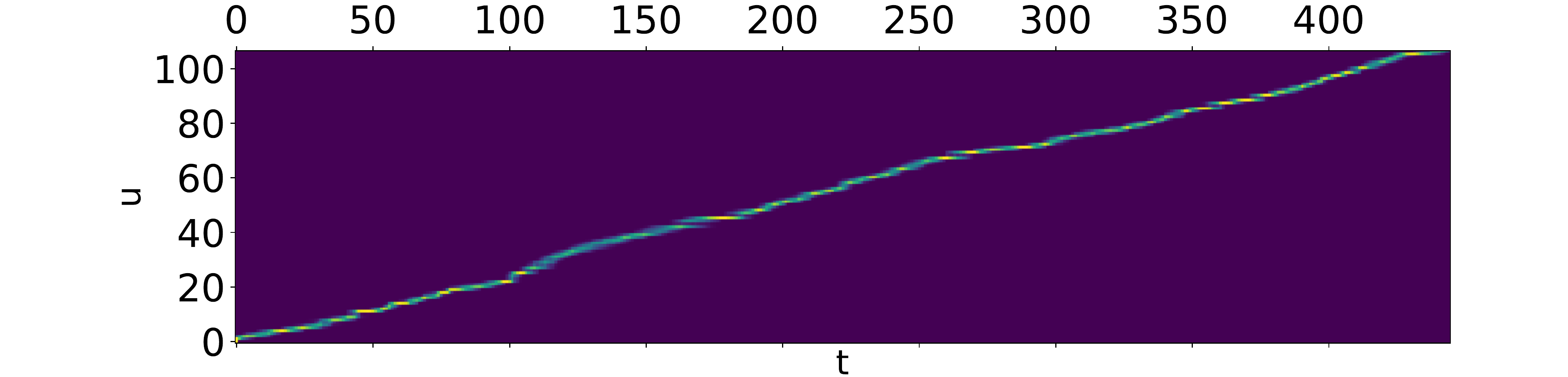}\\
\textbf{(b)}\\
\end{tabular}
\caption{Visualizations of node gradient in the standard transducer
lattice for an utterance during training. The joiner consists of an adder and
a softmax layer.
\textbf{(a)} The model parameters are randomly initialized.
\textbf{(b)} The model has been trained for several epochs.
}
   \label{fig:lattice-node-grad}
  \vspace{-2mm}
\end{figure}

Figure~\ref{fig:lattice-node-grad} shows the node gradient at each position
$(t,u)$ in Figure~\ref{fig:lattice}(a). It shows that: (1) At each time
step, there is only a small range of nodes with a non-zero gradient;
(2) Positions of nodes with non-zero gradient change monotonically from the lower
left to the upper right. Therefore, instead of generating a probability distribution
over all $U$ tokens at each time step, we can limit the range of tokens from
$U$ to $S$, where $S \ll U$. That is, at time step $t$, we only compute the
log-probabilities at positions $(t,p_t), (t,p_t+1), \ldots, (t, p_t+S-1)$ and
the log-probabilities at other positions are set to minus infinity.
Figure~\ref{fig:lattice}(b) shows the lattice for pruned RNN-T when
$S$ is 4.

As a consequence, the output shape of the joint network in the pruned RNN-T
becomes $(T, S, V)$. Since $S \ll U$, it reduces not only memory consumption
but also computation, thus leading to faster training.

%% file: pruned-rnnt.tex
 \section{Pruned RNN-T}
 \label{sec:pruned-rnnt}

 The basic idea of pruned RNN-T is to only evaluate the joiner network for
 the $(t, u)$ pairs that significantly contribute to the final loss.  We
 do this by computing the core recursion of~\eqref{eq:standard-forward} twice.
 The first time we do it with a ``trivial'' joiner network that is very fast
 to evaluate; we use this output to work out which indexes are important,
 and evaluate the full joiner network with a subset of $(t, u)$ pairs.

 \subsection{Trivial joiner network}

 We formulate the trivial joiner network in such a way that the computation of
 $y(t, u)$ and $\varnothing(t, u)$ can be implemented by matrix multiplication
 and some simple lookups.  (We omit the subscript $\mathrm{trivial}$ here for
 brevity; let it be understood that in this Sec.~\ref{sec:pruned-rnnt} we are talking about a separate version of
 these variables).  In the trivial version of the joiner we project the
 encoder embedding and the decoder embedding respectively to un-normalized
 logprobs  $L_\mathrm{enc}(t, v)$ and $L_\mathrm{dec}(t, u)$
 respectively, and the joiner consists of simply adding these together
 and normalizing the log-probabilities:
\begin{equation}
  L_\mathrm{trivial}(t, u, v) \eqdef L_\mathrm{enc}(t, v) + L_\mathrm{dec}(u, v) - L_\mathrm{normalizer}(t, u), \label{eqn:l_trivial}
\end{equation}
where
\begin{equation}
  L_\mathrm{normalizer}(t, u) \eqdef \log \sum_v \exp \left( L_\mathrm{enc}(t, v) + L_\mathrm{dec}(u, v) \right) . \label{eqn:l_normalize}
\end{equation}
Equation~\eqref{eqn:l_normalize} can be thought of as log-space matrix multiplication,
and it can be implemented by conventional matrix multiplication after first applying offsets
to ensure that overflow will not occur.

Thus, with the trivial joiner network we can compute $y(t, u)$ and $\varnothing(t, u)$ without
ever materializing any ``large'' matrices.

 \subsection{Pruning bounds}

 We introduce a constant $S$, e.g. $S=4$ or $S=5$, that represents the number of $u$
 indexes that we will evaluate for any $t$ index\footnote{This is called s\_range in the code.}.
 For each $t$ we will evaluate $L(t, u, v)$ only for integer positions $p_t \leq u < p_t + S$, where $p_t$
 (representing a position on the $u$ axis)  will be computed from the result of the recursion of the trivial joiner network.
 We will construct the $y(t, u)$ and $\varnothing(t, u)$ matrices with only these elements
 set, and all others set to $-\infty$, before doing the recursion~\eqref{eq:standard-forward}.

 \subsubsection{Globally optimal pruning bounds}

 Ideally, we would like to find the sequence of integer pruning bounds
 $p = p_0, p_t, \ldots, p_{T-1}$ that would maximize the total retained
 probability based on the ``trivial'' joiner network, treating all
 $P(t, u, \cdot) $ for $u < p_t$ or $u \ge p_t + S$ as $-\infty$.  That
 is, we are searching for the pruning bounds that would maximize the data likelihood given
 the trivial joiner, after setting all pruned logprobs to $-\infty$.
 This is a difficult optimization problem.  Instead, we solve it
 by finding the {\em locally optimal} pruning bounds for each frame $t$,
 and then applying some continuity constraints to the result.

 \subsubsection{Locally optimal pruning bounds}

 The total data log-prob is given by $L_\mathrm{tot} = \alpha(T-1,U) + \varnothing(T-1, U)$;
 let it be understood that we are talking about the ``trivial'' version of these
 variables.  Define $y'(t, u)$ and  $\varnothing'(t, u)$ as the derivatives of $L_\mathrm{tot}$ with respect to $y(t, u)$
 and $\varnothing(t, u)$.  We will need to compute these later anyway, in the neural network
 backprop; we do this ``early'' in the forward pass so that we can use the derivatives to
 compute the pruning bounds.

 You can think of $y'(t, u)$ and $\varnothing'(t, u)$ as ``occupation counts''
 in the interval $[0,1]$, which correspond to the probability of taking the upward
 and rightward transitions in Figure~\ref{fig:lattice}(a). Now consider the case where $S = 4$
 and we want to compute how much the total data log-probability would be decreased
 if we were to choose $p_t = 2$, for example.  The total amount of retained
 probability mass can be lower-bounded by
 \begin{equation}
   \varnothing'(t, 2) + \varnothing'(t, 3) + \varnothing'(t, 4) + \varnothing'(t, 5) - y'(t, 1),
 \end{equation}
 indicated by the colored transitions in Figure~\ref{fig:lattice}(a).
 Our ``locally-optimal'' $p_t$ is thus
 \begin{equation}
  p_t =  \operatorname{argmax}_{p=0}^{U{-}S{+}1}( -y'(t, p - 1) + \sum_{u=p}^{p{+}S{-}1} \varnothing'(t, u)) . \label{eqn:pt}
 \end{equation}
 \footnote{The experiments in this paper were actually done with an earlier, less-accurate version
   of this computation}  To explain~\eqref{eqn:pt}, which corresponds to ``green arcs minus red arc''
 in Figure~\ref{fig:lattice}(a): we want to get the total probability mass of
 all the paths that will be included if we use this pruning bound; and we can do by summing up
 the probabilities of exactly one link in each included path; \eqref{eqn:pt} is not the only
 way to do this.  The arcs summed in~\eqref{eqn:pt}
 include some probability mass that would actually be pruned out because it arises from
 lower $u$ values, and we cancel this by subtracting $y'(t, p - 1)$, which is red in the
 diagram. This may slightly over-compensate, to the extent that some of that subtracted
 probability mass makes it all the way through the pruned region.

 It is possible for~\eqref{eqn:pt} to give a sequence of $p_t$ values that are ``inconsistent'', i.e.
 that do not admit any complete path.  For consistency, in addition to $0 \leq p_t \leq U-S+1$, we
 require:
\begin{align}
  p_t           & \le p_{t+1} \label{eq:monotonic}\\
  p_{t+1} - p_t & < S \label{eq:no-skip} .
\end{align}
We modify the pruning bounds $p_t$ after computing them with~\eqref{eqn:pt} to ensure
that they satisfy these constraints while otherwise changing them as little as possible.
We won't expand due to length constraints\footnote{Search for \_adjust\_pruning\_lower\_bound in \url{https://github.com/k2-fsa/k2} for more details.}.

%%  We have implemented a primitive called $\operatorname{MonotonicLowerBound}(\cdot)$ that
%%  can be used to modify the sequence $p_t$.  For an arbitrary input sequence $a_i$, this
%% function returns the most-positive sequence of values $b_i$ such that $b_i \leq a_i$ and
%% $b_i \leq b_{i+1}$, equivalent to: $b_i = {\operatorname{min}}(a_i, a_{i+1}, \ldots)$.
%% We can enforce the constraints of~\eqref{eq:monotonic} and~\eqref{eq:no-skip} with
%% minimal changes to the sequence ${\b p} = p_t, 0 \leq t < T$, as follows
%% \begin{eqnarray}
%%   {\b p} &\leftarrow& \operatorname{MonotonicLowerBound}(\b p) \\
%%   p_t    &\leftarrow& -p_t  +  (S-1) t , \ \ \ \forall t   \\
%%   {\b p} &\leftarrow& \operatorname{MonotonicLowerBound}(\b p) \\
%%   p_t    &\leftarrow& -p_t  +  (S-1) t , \ \ \ \forall t \\
%%   p_t    &\leftarrow& max(0, p_t) , \ \ \ \forall t .
%% \end{eqnarray}
%% It is not clear that this is optimal in any sense; but these continuity
%% constraints are rarely active, only needed to avoid rare failures.

 \subsection{Loss function}

 The loss function will be a combination of the data log-probability from
 the trivial joiner and the one with the full joiner network.   If the full-joiner
 logprob is unscaled, we found it best to scale the trivial-joiner logprob
 by $0.5$ in the loss function; it seems to have a regularizing effect.

 \subsubsection{Smoothed trivial joiner}

 Since~\eqref{eqn:l_trivial} makes it natural to separate the encoder (acoustic) and decoder (language-model/LM) parts
 of $L_\mathrm{trivial}(t, u, v)$, we decided to try interpolating the trivial joiner with even-more-trivial versions
 of the joiner network: specifically, versions where we use encoder-only and decoder-only versions of the probabilities.
 Let $\operatorname{LogSoftmax}$ be the log-softmax operation, applied along the appropriate axis (the v axis).
 So the version of the log-likelihoods we use in the recursion would be:
 \begin{align}
   L_\mathrm{smoothed}(t, u, v) =& \left(1-\alpha^\mathrm{lm}-\alpha^\mathrm{acoustic}\right) L_\mathrm{trivial}(t, u, v) \nonumber\\
                                &+ \alpha^\mathrm{lm} L_\mathrm{lm}(t, u, v) \nonumber\\
                                &+ \alpha^\mathrm{acoustic} L_\mathrm{lm}(t, u, v)
 \end{align}
 where:
 % The equations are too long
 % \begin{eqnarray}
 %   L_\mathrm{trivial}(t, u, v) &\eqdef& \operatorname{LogSoftmax}_v \left( L_\mathrm{enc}(t, v) + L_\mathrm{dec}(u, v) \right)  \\
 %   L_\mathrm{enc}(t, u, v) &\eqdef& \operatorname{LogSoftmax}_v \left( L_\mathrm{enc}(t, v) + L_\mathrm{dec}^\mathrm{avg} \right)  \\
 %   L_\mathrm{dec}(t, u, v) &\eqdef& \operatorname{LogSoftmax}_v  L_\mathrm{dec}(u, v) .
 % \end{eqnarray}
 \begin{equation}
   L_\mathrm{trivial}(t, u, v) \eqdef \operatorname{LogSoftmax}_v \left( L_\mathrm{enc}(t, v) + L_\mathrm{dec}(u, v) \right)
 \end{equation}
 \begin{equation}
   L_\mathrm{acoustic}(t, u, v) \eqdef \operatorname{LogSoftmax}_v \left( L_\mathrm{enc}(t, v) + L_\mathrm{dec}^\mathrm{avg}(u,v) \right)  \\
 \end{equation}
 \begin{equation}
   L_\mathrm{lm}(t, u, v) \eqdef \operatorname{LogSoftmax}_v  L_\mathrm{dec}(u, v) .
 \end{equation}
 and $L_\mathrm{dec}^\mathrm{avg}(u, v)$ takes the role of a unigram language-model prior:
 \begin{equation}
   L_\mathrm{dec}^\mathrm{avg}(u, v)  \eqdef \log \frac{1}{U+1} \sum_{u=0}^{U} \operatorname{Softmax}_v L_\mathrm{dec}(u, v)
 \end{equation}
 The reason for the asymmetry between the encoder and decoder here is that we want the decoder log-probs
 to be independently interpretable as language-model probabilities; this will be more convenient in case
 we need to access the language model probabilities independently for some reason later on.

%% file: experiments.tex
\section{Experiment Settings}
\label{sec:experiments}

The experiment contains 2 parts. In the first part, we benchmark the speed
and memory usage of pruned RNN-T and several other open-source implementations
for computing RNN-T loss. In the second part, we apply pruned RNN-T for ASR
training using the LibriSpeech corpus~\cite{librispeech2015}. Note that we
don't use any kind of language models during decoding.

\subsection{Benchmarks of RNN-T loss computation}
We compare the speed and peak memory usage of pruned RNN-T
with the following open-source implementations:
warp-transducer\footnote{\url{https://github.com/b-flo/warp-transducer/tree/espnet\_v1.1}},
torchaudio~\cite{yang2021torchaudio},
optimized\_transducer\footnote{\url{https://github.com/csukuangfj/optimized\_transducer}},
and SpeechBrain~\cite{speechbrain}.

Because padding matters in the RNN-T loss computation and to make the benchmark
more realistic, instead of generating random data with random shapes we get the
shapes for tokens and acoustics using the test-clean dataset from the LibriSpeech
corpus. Two commonly used settings are benchmarked:
(1) Fixed batch size. In this setting, the batch size is fixed and
utterances in a batch have various lengths of durations.
(2) Dynamic batch size. In this setting, we sort utterances by durations before
batching them up to minimize paddings and the maximum number of frames
in a batch before padding is limited to 10k at 100 frames per second.

The number of model output units is 500.
We use an NVIDIA V100 GPU with 32 GB RAM to run the benchmarks.
The code is open-sourced and publicly
available\footnote{https://github.com/csukuangfj/transducer-loss-benchmarking}.

\subsection{Pruned RNN-T for ASR training}

We use pruned RNN-T for ASR training with the LibriSpeech corpus, which
consists of 960 hours of 16 kHz read English speech for training and
two subsets, test-clean and test-other, for testing, each of which has
approximately 5 hours speech data.

The inputs of the neural network model are 80-dimension log Mel filter bank
features with a window size 25 ms and a window shift 10 ms.
SpecAugment~\cite{park2019specaug} and speed perturbation~\cite{ko2015audio}
with factors 0.9 and 1.1 are used to make the training more stable. The outputs
of the model are 500 sentence pieces~\cite{kudo-richardson-2018-sentencepiece}
with byte pair encoding (BPE)~\cite{sennrich2016neural}.

The encoder of the RNN-T model is a Conformer~\cite{anmol2020conformer}
with 12 layers. Each encoder layer has 8 self-attention\cite{vaswani2017attention} heads.
The attention dimension and the feed-forward dimension are 512 and 2048,
respectively.
We use a stateless decoder~\cite{ghodsi2020rnntstateless}, which consists of
an embedding layer followed by a 1-D convolutional layer with a kernel size 2.
The embedding dimension is 512.
We use 8 NVIDIA V100 32GB GPUs for training.

To ensure
convergence, for the first few thousand minibatches we disable the pruned part
of the loss by giving it a zero scale in the loss function; after the trivial
loss starts to learn something meaningful, it will provide reasonable pruning
bounds and we can enable the pruned loss.   Due to space constraints we are not
showing convergence results, but the pruned
transducer converges very similarly to the conventional transducer.

%%The code is open-sourced and publicly
%%available\footnote{https://github.com/k2-fsa/icefall}.

%% file: results.tex
\section{Results}
\label{sec:results}

\subsection{Benchmark results}

\begin{table}[t]
\caption{Speed and memory usage for different RNN-T loss implementations
using fixed batch size 30.}\label{tbl:fixed-batch}
\begin{center}
\vspace{-10pt}
\begin{tabular}{lrr}
\hline
& Average time & Peak memory \\
&  per batch (ms) &  usage (GB) \\\hline
torchaudio & 544 & 18.48\\%18921.8/1024 = 18.47832
optimized\_transducer & 377 & 7.32\\%7495.9/1024=7.32
warp-transducer& 276 & 18.63 \\%19072.6/1024=18.625
SpeechBrain & 459 &  18.63\\%19072.8/1024=18.625
pruned RNN-T & \textbf{64} & \textbf{3.73}\\\hline%3820.3/1024=3.73
\end{tabular}
\end{center}
\vspace{-5pt}
\end{table}

\begin{table}[t]
\caption{Speed and memory usage for different RNN-T loss implementations
using dynamic batch size where utterances are sorted by durations
before batching them up and the maximum number of frames in a batch
before padding is limited to 10k at 100 frames per second.}\label{tbl:dynamic-batch}
\vspace{-10pt}
\begin{center}
\begin{tabular}{lrr}
\hline
& Average time & Peak memory \\
&  per batch (ms) &  usage (GB) \\\hline
torchaudio & 601 & 12.66\\%12959.2/1024 = 12.655
optimized\_transducer & 568 & 10.65\\%10903.1/1024=10.647
warp-transducer& 211 & 12.76 \\%13061.8.6/1024=12.755
SpeechBrain & 264 &  12.76\\%13063.4/1024=12.757
pruned RNN-T & \textbf{38} & \textbf{2.59}\\\hline%2647.8/1024=2.585
\end{tabular}
\end{center}
% \vspace{-25pt}
\vspace{-5pt}
\end{table}

\begin{table}[!h]
\caption{WERs on the LibriSpeech test-clean and test-other datasets
for two models, where one is trained with pruned RNN-T and the other
is trained using optimized\_transducer.
Beam search with beam size 4 is used for decoding. No external LMs
are used during decoding.}\label{tbl:wer-comparison}
\vspace{-10pt}
\begin{center}
\begin{tabular}{lrrr}
\hline
\multirow{2}{*}{} & \multicolumn{2}{c}{test} & train hours\\
& clean & other & per epoch\\\hline
pruned RNN-T & 2.56 & 6.27 & 1.17\\
optimized\_transducer & 2.61 & 6.46 & 2.33\\\hline
\end{tabular}
\end{center}
\vspace{-10pt}
\end{table}

Table~\ref{tbl:fixed-batch} and Table~\ref{tbl:dynamic-batch} compare the
speed and peak memory usage for different implementations.
Pruned RNN-T has a clear advantage in not only speed but also memory usage in both benchmark settings.
The memory efficiency means that we can use a larger batch size and vocabulary size during training,
further increasing speed.

\subsection{Results for ASR training}

Table \ref{tbl:wer-comparison} compares WERs on the LibriSpeech
test-clean and test-other datasets for models trained
with pruned RNN-T vs. optimized\_transducer.
Pruned RNN-T has slightly better WER than the model trained with
unpruned RNN-T loss.

%% file: conclusion.tex
\section{Conclusions}
\label{sec:conclusions}
In this paper, we propose pruned RNN-T to reduce the memory usage
and computation for RNN-T loss by limiting the range of tokens
at each time step. Benchmark results show that pruned RNN-T is the fastest
and consumes the least memory among commonly used open-source implementations:
torchaudio, warp-transducer, SpeechBrain, and optimized\_transducer.
Furthermore, we demonstrate that using pruned RNN-T in ASR training can
achieve competitive WERs with standard RNN-T on the LibriSpeech corpus.